\documentclass[a4paper]{easychair}

\usepackage{doc}
\usepackage{makeidx}

\usepackage{booktabs}
\usepackage{graphicx}

\usepackage{subcaption}
\usepackage{epstopdf} 

\usepackage{listings}

\title{Using Cognitive Computing for Learning Parallel Programming: An IBM Watson Solution}

\titlerunning{Using Cognitive Computing for Learning Parallel Programming}

\author{
    Adri\'{a}n Calvo Chozas 
\and
    Suejb Memeti 
\and
    Sabri Pllana 
}

\institute{
  Linnaeus University,
  V\"{a}xj\"{o} 351 95, Sweden\\
  \email{ac222ne@student.lnu.se}\\
  \email{suejb.memeti@lnu.se}\\
  \email{sabri.pllana@lnu.se}
 }

\authorrunning{A. C. Chazos, S.Memeti, and S. Pllana}

\begin{document}

\maketitle

\begin{abstract}
While modern parallel computing systems provide high performance resources, utilizing them to the highest extent requires advanced programming expertise. Programming for parallel computing systems is much more difficult than programming for sequential systems. OpenMP is an extension of C++ programming language that enables to express parallelism using compiler directives. While OpenMP alleviates parallel programming by reducing the lines of code that the programmer needs to write, deciding how and when to use these compiler directives is up to the programmer. Novice programmers may make mistakes that may lead to performance degradation or unexpected program behavior. Cognitive computing has shown impressive results in various domains, such as health or marketing. In this paper, we describe the use of IBM Watson cognitive system for education of novice parallel programmers. Using the dialogue service of the IBM Watson we have developed a solution that assists the programmer in avoiding common OpenMP mistakes. To evaluate our approach we have conducted a survey with a number of novice parallel programmers at the Linnaeus University, and obtained encouraging results with respect to usefulness of our approach.
\end{abstract}

\section{Introduction}
\label{sect:introduction}

The modern parallel computing systems provide capabilities to solve complex computational and engineering problems \cite{shiflet2011introduction} faster. Programming for parallel computing systems is much more complex than programming for sequential processors \cite{sarkar2007,pllana2008}, because it requires knowledge of the underlying parallel architecture and the programming models libraries (that are often device specific), and consideration of large amount of device specific configuration parameters (such as, numbers of cores, core speed, memory hierarchy level, cache, run-time system, etc) \cite{sandrieser12}. 

To alleviate the programmability challenges of parallel computing systems \cite{Kessler12,benkner11,viebke2017}, various parallel programming models and languages \cite{parasurvey} are proposed, including OpenMP, OpenACC, MPI, Pthreads, OpenCL, Cuda, and Intel TBB. While these approaches have helped to significantly reduce the programming effort, writing more complex code requires more effort, advanced knowledge of parallel algorithms and underlying architecture and is more prone to mistakes that may lead to incorrect program behavior or performance degradation \cite{goncalves2015}. To efficiently utilize the available resources on the modern parallel computing systems, programmers require adequate education \cite{czarnul2014}.

OpenMP (Open Multi-Processing) is an Application Program Interface (API) that comprises a set of compiler directives, variables and library functions to program parallel computing systems. OpenMP is implemented as language extension of C, C++ and Fortran \cite{dagum1998} for shared-memory parallel computing systems, and recent versions of OpenMP support also the heterogeneous computing systems. The main goal of OpenMP is to reduce the difficulty of writing parallel applications. It does that by using compiler directives, known as pragmas. However, it is up to the programmer to decide how and when to use these directives. Even though, the available documentation suggests that using OpenMP is simple and easy, Gon\c{c}alves et al. \cite{goncalves2015} show that using OpenMP is not as easy as it appears, and such suggestions may lead to code that provides incorrect results or poor performance \cite{perfmod,Pllana2005}. Furthermore, Kolosov et al. \cite{kolosov2008} and S{\"u}{\ss} et al. \cite{suss2008} has identified a number of commonly made OpenMP mistakes, and they have categorized them in logical and performance related errors.

Cognitive systems \cite{modha2011} are on the verge of becoming a milestone due to their ability to process natural language processing (NLP). This enables a whole new level of interaction between humans and computing systems, which may help people make better decisions. IBM Watson \cite{ferrucci2012} is a cognitive system that uses the natural language to receive and answer questions. It is able to handle unstructured data, which composes 80\% of the data available on the Web. IBM Watson learns a new subject by storing all the documents related to the topic on its own database, in which it later searches for the most suitable answer. Recently various solutions based on the IBM Watson have demonstrated impressive results \cite{computerworld2017}. For example, in 2011 the Watson Dialog service won at Jeopardy against two human former winners. Then, in 2012 Watson helped oncologists to extract valuable information to treat patients. \emph{Standard Bank} is using Watson to solve customers' queries. The \emph{American Cancer Society} is using Watson to provide better answers to patients. 

In this paper, we describe the use of cognitive computing to aid novice programmers learn parallel programming. We have developed an application based on the IBM Watson services that enables a dialog-based interaction with programmers during program development. Our solution is focused on helping novice programmers to avoid commonly made mistakes when they use OpenMP. We have implemented our solution to support interaction with the programmer in English and Spanish. Furthermore, we have conducted a survey to evaluate the usability of our solution. The results of the survey indicate that the developed application offers valuable answers that would be helpful for novice parallel programmers. 

The key contributions of this paper are:

\begin{itemize}
	\item A multi-language IBM Watson application that aids novice programmers to learn parallel programming with OpenMP,
	\item Evaluation of usability of our IBM Watson application.
\end{itemize}

The rest of the paper is organized as follows. In Section \ref{sect:rw} we discuss the related work. Section \ref{sect:methodology} describes the methodology and implementation details. In Section \ref{sect:evaluation} we first describe the context of the conducted survey, thereafter we discuss the results of the survey. We conclude our paper and provide future research directions in Section \ref{sect:conclusion-future-work}.

\section{Related Work}
\label{sect:rw}

Goel et al. \cite{goel2015} used different Watson services to develop six diverse applications that aim at understanding the functionality and capabilities of Watson and enhancing the human-computer co-creativity. By developing these diverse applications the authors argue that Watson has potential to be used in different domains, and has large range of opportunities to be used as an educational tool.

Witte et al. \cite{witte2011} uses natural language processing to bring new levels of support to software developers. A plug-in that is integrated in the Eclipse IDE is developed, which provides quality analysis of the comment found in source code and version control commits. The aim of this project is to help software developers reduce the effort required to analyze their code by extracting useful information that might be valuable to understand the functionality of the application that is not always obvious by looking at the source code only.  

An interactive tool for parallelization of OpenMP sequential programs is proposed by Ishihara et al. \cite{ishihara2003}. This tool provide four types of assistance to the developers, including analysis of the parallelism, identification of candidate parallel sections, enhancing the parallelism through code restructuring, and analysis of the execution time. 

Harms \cite{harms2014} proposes an approach that is able to monitor and understand the programming skill level of the developer and adaptively suggest code examples that may help to learn new programming concepts found within the suggested examples. The author argues that this approach avoid overwhelming the memory of novice programmers by considering the previous knowledge of the programmer and carefully suggesting examples that contain new information. 

Sah and Vaidya \cite{sah2012} review, discuss characteristics and classify existing parallelization tools based on: (1) the time the tools were developed (before or after the multicore era); (2) their contribution during the parallelization process, and (3) the ability to provide assistance for parallel programming. Furthermore, they propose their parallel programming tool that assist developers during the software development time. 

In contrast to the related work, we use cognitive computing to assist programmers in avoiding common logical and performance mistakes. Our solution is based on the IBM Watson dialogue service and focuses on OpenMP.

\section{Our methodology}
\label{sect:methodology}

In this section first we will describe the design of our approach thereafter we will provide some implementation details.

\subsection{Design}

\begin{figure}[tb]
	\centering
	\includegraphics[width=0.55\linewidth]{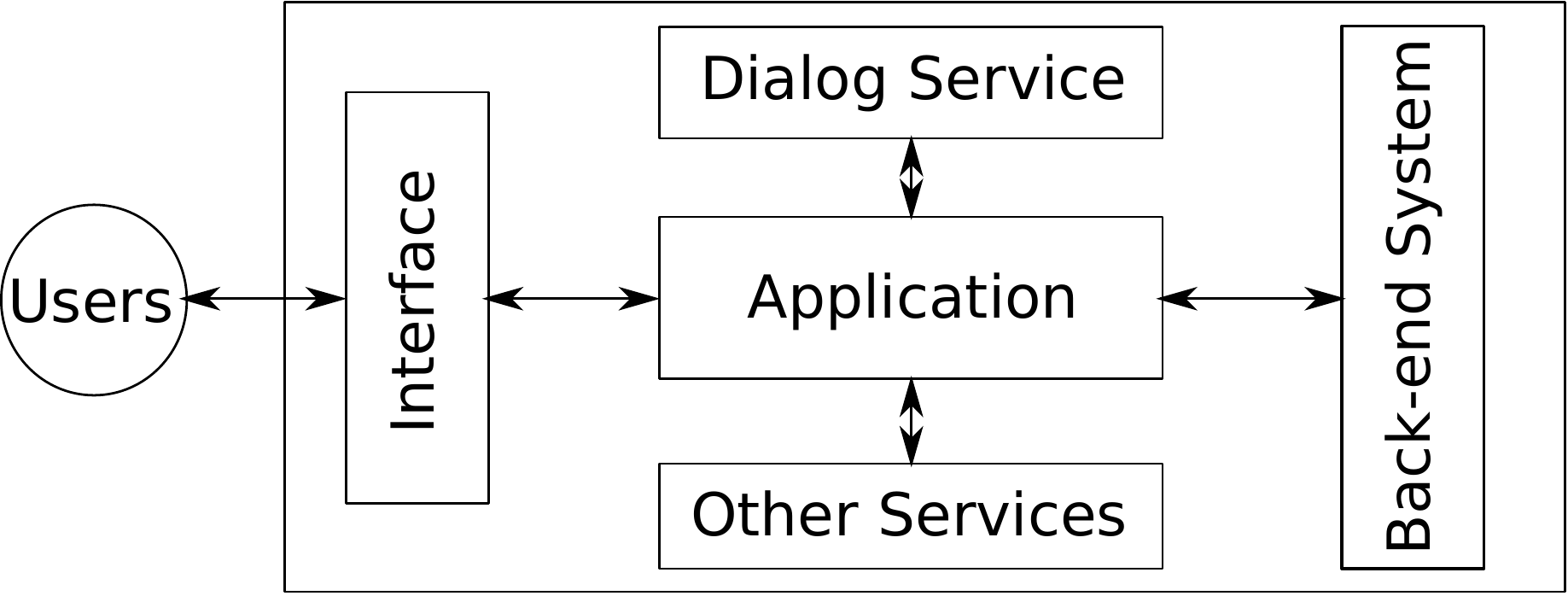}
	\caption{Architecture overview}
	\label{fig:application-overview}
\end{figure}

Our application is based on the IBM Watson Dialog service to provide communication means between the user and the system through natural language processing with the aim to avoid common OpenMP mistakes. The architecture overview of our application is depicted in Figure \ref{fig:application-overview}. The users interact with our application using an interface, which allows them to write questions in text form. The user that will presumably be a parallel programmer will use the interface to ask a question related to parallel programming. Please note that as is usual in natural language, the question can be formulated in several ways. For accessibility, the interface may be enhanced with voice recognition or connected to some kind of integrated development environment. 

The interface sends the user input to our application, which is connected to the Watson Dialog Service. The dialog service analyzes the questions and gathers information that it needs to provide the most suitable answer for the user. To extend the functionality of our application, one can connect the application to additional services such as speech-to-text services. 

The application is connected to a back-end system that is responsible to display the answer of the question to the user. At this point, the application has  the  answer for the user. There are two scenarios, (1) the  question  has  been understood and application provides a proper answer to the user, and (2) the question has not been understood and a default  message saying \textquotedblleft \emph{I  am  sorry,  I  did  not  understand  your  question. Please try another question}.\textquotedblright is provided to the user.
 
\begin{figure}[bt]
	\centering
	\includegraphics[width=0.55\linewidth]{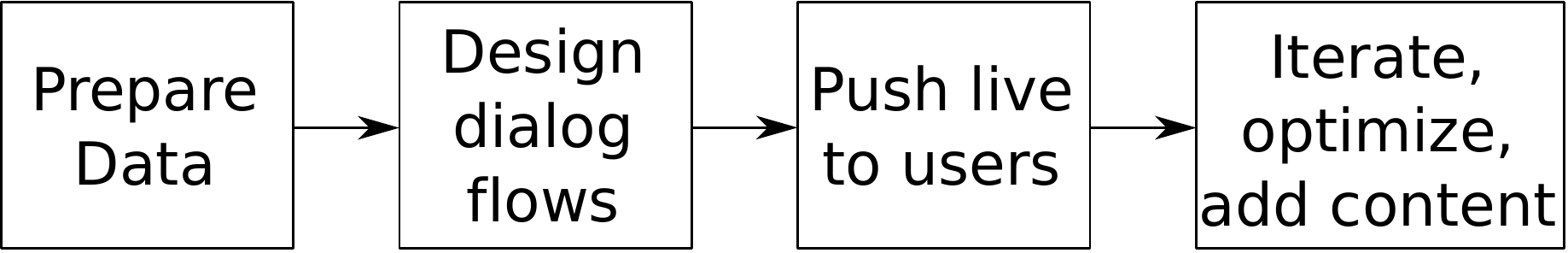}
	\caption{An overview of the Watson Dialog Service}
	\label{fig:dialog-service}
\end{figure}

The overview of the IBM Watson dialog service is illustrated in Figure \ref{fig:dialog-service}. The dialog service provides means for communication between computers and the users in a question-and-answer fashion through natural language. To use this service, first the training data should be gathered and prepared by an expert. As our aim is to provide a tool that aids novice parallel programmers write parallel code using OpenMP, we considered to use as training data the common OpenMP mistakes identified by Kolosov et al. \cite{kolosov2008} and S{\"u}{\ss} et al. \cite{suss2008}. By the time of writing this paper, there are 32 commonly identified mistakes \footnote{The list of all 32 identified OpenMP mistakes can be found in: https://software.intel.com/en-us/articles/32-openmp-traps-for-c-developers}, however due to space limitations we have listed only some of them in Table \ref{tab:performance-logical-mistakes}.

\begin{table}[tb]
	\centering
	\footnotesize
	\caption{Common OpenMP mistakes identified by Kolosov et al. \cite{kolosov2008} and S{\"u}{\ss} et al. \cite{suss2008}.}
	\label{tab:performance-logical-mistakes}
	\begin{tabular}{@{}p{5cm}p{9cm}@{}}
		\toprule
		Performance errors      			& 	Reason \\ \midrule
		Unnecessary flush					&	If flush directive is used without parameters, it can reduce the performance of the program. \\
		Using critical instead of atomic	&	The atomic directive is faster than critical. When atomic cannot be used, the compiler will not allow the programmer to use it. \\
		Unnecessary protection from concurrent memory write & Local thread variables should not be protected from concurrent writing \\
		Overwork in a critical region		&	It is known that critical regions reduce the  performance  of  the  program   so using critical is generally not recommended.\\
		\toprule
		Logical errors                    	& Reason\\ \midrule
		Missing \textit{/openmp}          	& If OpenMP is not enabled in the compiler settings, the OpenMP directives will be ignored \\
		Missing \textit{parallel}  		  	& If the programmer forgets to put the parallel keyword, the code will run sequentially \\
		Missing \textit{omp}              	& If omp keyword is forgotten the entire pragma will be ignored \\
		Missing \textit{for}              	& If the programmer want to divide a loop into $n$ threads and \textit{for} is forgotten, the program will not split up the work into those n threads \\
		Unnecessary parallelization		  	& If the programmer puts the parallel keyword inside a parallel section, the loop will be run $n$ times \\
		Incorrect usage of \textit{ordered} & If the ordered directive is not correctly indicated, the compiler will decide to order randomly \\ 
		Redefining the number of threads in a parallel section & Attempts to change the number of threads within a parallel region will result with run-time errors. \\
		Lock a variable without initializing it & According to the specification, to lock a variable it first needs to be initialized \\
		Unsetting locks from another thread & Locks set from one thread will cause unpredictable run-time behavior if unset from another thread \\
		Parallel array without order & If the result depends on previous iterations, order clause is compulsory in order for it not to have unexpected behaviour. \\
		Access to a share memory without protection & When several threads are modifying a variable the result is unpredictable.\\		
		\bottomrule
	\end{tabular}
\end{table}

\begin{figure}[bt]
	\centering
	\begin{subfigure}[b]{0.48\textwidth}
		\includegraphics[width=\textwidth]{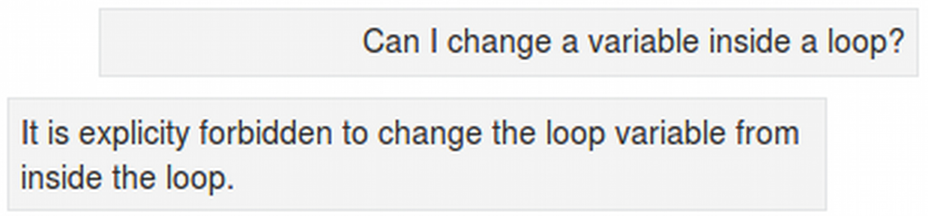}
		\caption{English}
		\label{fig:english}
	\end{subfigure}
	\hfill
	~ 
	\begin{subfigure}[b]{0.48\textwidth}
		\includegraphics[width=\textwidth]{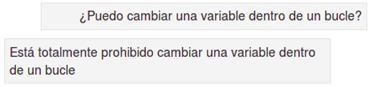}
		\caption{Spanish}
		\label{fig:spanish}
		
	\end{subfigure}
	
\caption{Example of interaction with the programmer in English and Spanish}
	\label{fig:multilang}
\end{figure}

The gathered data should be translated into dialog models that is part of the dialog flow design process. For instance, the common OpenMP mistake where programmers redefine the number of threads within a parallel loop, will be translated in a dialog fashion where the input would be like: \emph{Can I change a variable inside a pragma omp loop?}. Figure~\ref{fig:multilang} depicts the interaction with the programmer in English and Spanish.

The push live to users enables the connection of the dialog service with the user interface, and monitoring the conversations. The last component enables dynamic learning from real interactions by adjusting the existing content and adding new one based on the user activity. 

\subsection{Implementation Details}

In order to build an application that uses IBM Watson Dialog service the developer needs to specify the mandatory settings for the service as shown in the Listing \ref{lst:settings}. For example, the \emph{AUTOLEARN} (see line 4) setting allows the dialog to suggest another node when the input does not match. The \emph{LANGUAGE} (see line 3) setting allows the user to determine the account language. Please note that we show only an excerpt of all possible settings, for the full list of available setting we refer the reader to the Watson Dialog Service documentation \cite{watsondocumentation2017}.

\lstset{caption={An example of settings configuration of the Dialog Service},label=lst:settings,numbers=right,numbersep=-10pt,basicstyle=\footnotesize}
\begin{lstlisting}
<settings>
  <setting name="DISPLAYNAME" type="USER">test</setting>
  <setting name="LANGUAGE" type="USER">EN</setting>
  <setting name="AUTOLEARN" type"USER">false</setting>
  ...
</settings>  
\end{lstlisting}

The main characteristic of the Dialog service is the \emph{dialog} tag, which has a mandatory child named \emph{flow} that is responsible for the content of the dialog because it includes the main libraries (see Listing \ref{lst:dialog}). The \emph{flow} tag may have a child named \emph{folder} that is used to allocate, organize and maintain the information in the dialog. There are four types of folders, \emph{Main, Library, Global, and Concepts}.

\lstset{caption={An example of the dialog tags},label=lst:dialog,numbers=right,numbersep=-10pt,basicstyle=\footnotesize}
\begin{lstlisting}
<dialog ... xsi:noNamespaceSchemaLocation="WatsonDialogDocument_1.0.xsd">
  <flow>
    <folder label="Main">
      ...
\end{lstlisting}

The \emph{Main} folder stores the welcome messages, whereas the \emph{Library} folder stores the core of the dialog such as, the input (an example of the input node is shown in Listing \ref{lst:input-node}), output, structures, and other valuable information. Each question is represented as an input node of the dialog service, where each node has a child grammar node, and the grammar node must have at least one child item node (see Listing \ref{lst:input-node}). As the dialog service is based on natural language, each question may have multiple variations. We can configure these variations by adding multiple items inside the grammar. The first item represents the primary variation of the question. We can have multiple variations by using the wildcards $\$$ (dollar sign) and $*$ (asterisk). 
When the dollar sign is used the question becomes more open as the user can type down the question in multiple ways, however the question must be followed by the first verb after the dollar sign (see Line 4 Listing \ref{lst:input-node}). The questions could be: \textquotedblleft \emph{\textbf{Is it possible} to change a variable inside a loop?}\textquotedblright, or  \textquotedblleft \emph{\textbf{May I} to change a variable inside a loop?}\textquotedblright. 
When the asterisk is in use it means that the place corresponding to the  asterisk  can  be  changed  for  another  suitable  word (see Line 5 Listing \ref{lst:input-node}), for example: \textquotedblleft \emph{Can I change \textbf{the number of} threads ...?}\textquotedblright, or \textquotedblleft \emph{Is it possible to change \textbf{all of the} threads ...?}\textquotedblright.

\lstset{language=C,caption={An example of the input node of the IBM Watson Dialog Service},label=lst:input-node,numbers=right,numbersep=-10pt,basicstyle=\footnotesize}
\begin{lstlisting}
<input>
  <grammar>
    <item>Can I change a variable inside a pragma omp loop?</item>
    <item>\$ Change a variable inside a loop?</item>
    <item>change * variable * loop</item>
  </grammar>
</input>
\end{lstlisting}

The \emph{output} tag is a child of the \emph{input} and is used to store the answers of the corresponding question.For example, the corresponding answer to the question variations represented in the input node in Listing \ref{lst:input-node} would be \textquotedblleft \emph{It is explicitly forbidden to change the variable from inside the loop.}\textquotedblright, and the corresponding xml node is shown in Listing \ref{lst:output}

\lstset{caption={An example of the output tags},label=lst:output,numbers=right,numbersep=-10pt,basicstyle=\footnotesize}
\begin{lstlisting}
<output>
  <prompt selectionType="RANDOM">
    <item>It is explicitly forbidden to change the loop variable ...</item>
  </prompt>
</output>
\end{lstlisting}

If the system does not understand the question or it does not have a proper answer, the system will display a default message that could be customized using the \emph{default} node (see Listing \ref{lst:default}). 

\lstset{caption={An example of the system's default answer when no suitable response is found},label=lst:default,numbers=right,numbersep=-10pt,basicstyle=\footnotesize}
\begin{lstlisting}
<default>
  <output>
    <prompt selectionType="RANDOM">
      <item>I did not understand the question. Please try again.</item>
    </prompt>
  </output>
</default>
\end{lstlisting}

Please note that from September 2016 IBM has retired the Dialog service, which disallows creation of new instances of this service. However, existing instances will continue to work until August 2017. An equivalent service, so called Conversation service has been introduced, and developers are encouraged to migrate their Dialog service based applications to the Conversation service.

\section{Evaluation}
\label{sect:evaluation}

To evaluate the usefulness of our approach we have conducted an anonymous survey within the Linnaeus University campus. The target group that we have surveyed includes bachelor students from the department of computer science that have little knowledge in parallel programming. In total, eight novice parallel programmers have accepted to take part on the survey. The process of selection of candidates, data collection and analysis, and reporting is adapted from the guidelines presented by Runeson and H\"{o}st \cite{runeson2009}. 

The aim of this survey is to investigate whether such interactive tools would help novice programmers solve parallel programming problems as well as identify ways to improve our tool. To keep the survey simple, and to avoid survey fatigue of our respondents, which usually happens because surveys are too long and include non-relevant questions, we have defined a survey with only four questions. Table \ref{tab:survey-questions} lists the questions of our survey.

To measure the performance quality of our tool we have used the Likert \cite{allen2007} scale system to allow respondents to rank the quality of our tool, where participants could choose answers between one and 5 stars, where \emph{One star} stands for \emph{Strongly disagree}, \emph{Two stars} stands for \emph{Somewhat disagree}, \emph{Three stars} means \emph{Neither agree nor disagree}, \emph{Four stars} means \emph{Somewhat agree}, and \emph{Five stars} stands for \emph{Strongly agree}.

\begin{figure}[bt]
	\centering
	\includegraphics[width=0.8\linewidth]{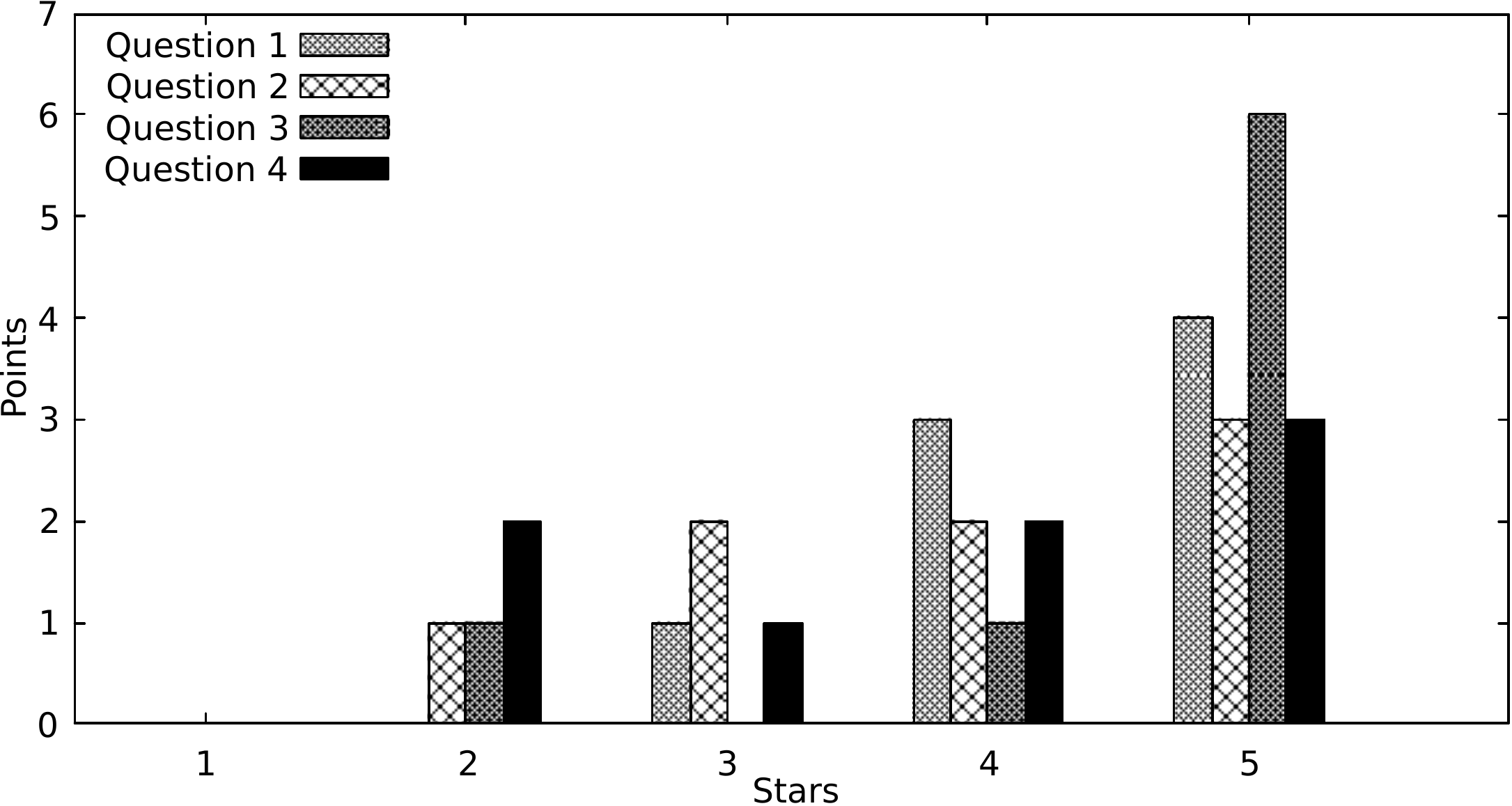}
	\caption{The results of the survey. The x-axis indicates the stars, whereas the y-axis indicates the number of participants that rated a question with a specific number of stars.}
	\label{fig:survey-results}
\end{figure}

The results of the survey are depicted in Figure \ref{fig:survey-results}. The x-axis indicates the stars, whereas the y-axis indicates the collected points such that if question 4 received 3 points for 5 stars it means that 3 participants strongly agreed that it would be useful to enhance our tool with the ability to retrieve papers as an answer. We may observe that most of the questions received more than three stars. For instance, when asked whether our tool is more helpful than using search engines or paper-based resources, 3 participants neither agreed nor disagreed, three other participants agreed to some extent, and four of them strongly agreed. When asked whether the answers retrieved from our tool were accurate, one of the participants disagreed to some extent, two participant were neutral, two participants agreed to some extent, and three of them strongly agreed. Six of the participants strongly agreed, one agreed to some extent and another one disagreed to some extent that our tool is useful because it can respond to questions asked in different languages. Two participants disagreed to some extent and two others agreed to some extent, another one was neutral, and three others strongly agreed that enhancing our tool with the ability to retrieve paper as an answer would be useful.

\begin{table}[tb]
	\centering
	\caption{The questions of the conducted survey}
	\label{tab:survey-questions}
	\begin{tabular}{@{}lp{13.5cm}@{}}
		\toprule
		\# 	& 	Question \\ \midrule
		1	&	Do you think that such interactive assistance application can help you in parallel programming, more than search engines or paper-based resources? \\
		2	&	Do you think that the answers you have received from our tool were accurate? \\
		3	& 	Do you find it useful that the application is multilingual? \\
		4	&	Would you find it useful if the application could retrieve papers as an answer?\\
		\bottomrule
	\end{tabular}
\end{table}

Table \ref{tab:survey-results} shows a matrix based representation of the percentage of each question related to the respondents' answers. The results show that the application would be successful among students as they would like to have an application that can help them in parallel programming by using natural language. In the first question 87.5\% of the total score corresponds to 4 stars or more, which means that the respondents highly agree with the usefulness of the application. In the second question 62.5\% of the total score corresponds to 4 stars or more, which means that the respondents highly agree with the usefulness of the application. In the third question 87.5\% of the total participants appreciated having multi-language support in this application. In the last question, 62.5\% of the users think that enhancing our tool with the ability to retrieve papers as answers is useful.

\begin{table}[tb]
	\centering
	\caption{ matrix based representation of the survey results that shows the percentage of each question related to the answers of the participants}
	\label{tab:survey-results}
	\begin{tabular}{@{}l|lllll@{}}
		\toprule
		& 1-star & 2-stars & 3-stars & 4-stars & 5-stars \\ \midrule
		Question 1 & 0\%    & 0\%     & 12.5\%  & 37.5\%  & 50\%    \\
		Question 2 & 0\%    & 12.5\%  & 25\%    & 25\%    & 37.5\%  \\
		Question 3 & 0\%    & 12.5\%  & 0\%     & 12.5\%  & 75\%    \\
		Question 4 & 0\%    & 25\%    & 12.5\%  & 25\%    & 37.5\%  \\ \bottomrule
	\end{tabular}
\end{table}

\section{Conclusion and Future Work}
\label{sect:conclusion-future-work}

Cognitive systems, like IBM Watson, are able to learn trough user's input and be taught by experts. An application that uses this technology can become a milestone in the field in which it is being used due to the fact that a community can be created to improve applications, share new ideas and discover features that no one has thought of before. The usefulness of such systems has been demonstrated in various fields, such as in health care, where doctors can treat their patients better owing to previous experiences with common diseases.

In this paper we have investigated the use of cognitive computing to help programmers learn parallel programming. We have used the IBM Watson dialog service to enable the interaction between the user and the system in a dialog-fashion way using natural language. Our application is trained to respond to questions related to common mistakes that novice programmers do when using OpenMP, which helps them avoid such mistakes, write code that produces the correct result and runs faster. Furthermore, such systems may reduce the time investment required to learn parallel programming. Since  Watson can learn from the user’s collected input, the application can be improved as it is used. The proposed tool can answer questions that a user may ask, but it will neither generate code nor find code errors.

To evaluate the usefulness of our application, we conducted a survey with a number of novice parallel programmers. The results of the survey show that novice parallel programmers are willing to use such interactive application while writing their parallel programs and would find it useful to improve their knowledge of the subject. Furthermore, the results show that retrieving papers as answers will significantly attract users to use our application. We believe that our tool can be used as an educational resource in a beginners course in parallel programming. 

Future research will address the extension of our solution with additional IBM Watson services that enable automatic input of data from existing sources of parallel programming knowledge.


\label{sect:bib}

\end{document}